\documentclass[12pt]{article}

\newtheorem{theorem}{Theorem}

\newtheorem{definition}[theorem]{Definition}

\newtheorem{proposition}[theorem]{Proposition}

\newtheorem{remark}[theorem]{Remark}

\RequirePackage[latin1]{inputenc}    
\usepackage[english]{babel}
\usepackage{amsmath}
\usepackage{amsfonts}
\usepackage{amssymb}
\usepackage{xspace}
\usepackage{latexsym}
\usepackage{url}
\usepackage{xspace}
\usepackage{fancyvrb}

\newcommand{\infer}[2]{\begin{array}{c} #1 \\ \hline #2 \end{array}}
\newcommand{\Eval}[1]{\Downarrow^{#1}}
\newcommand{\s}[1]{{\sf #1}}    
\newcommand{\w}[1]{{\it #1}}    
\newcommand{\pset}[1]{\{\! | #1 |\!\}}  
\newcommand{\vc}[1]{{\bf #1}}
\newcommand{\act}[1]{\stackrel{#1}{\rightarrow}} 
\newcommand{\union}{\cup}               
\newcommand{\Alt}{ \mid\!\!\mid  }
\newcommand{\lf}{\lfloor}
\newcommand{\rf}{\rfloor} 
\newcommand{\qqs}[2]{\forall\, #1\;\: #2}
\newcommand{\eval}{\Downarrow}
\newcommand{\xst}[2]{\exists\, #1\;\: #2}
\newcommand{\minus}{\backslash}         
\newcommand{\set}[1]{\{#1\}}            
\def\nil{(\!)}

\begin{document}


\title{Reactive concurrent programming revisited}

\author{Roberto M. Amadio
\thanks{Universit\'e Paris 7} 
\and
G\'erard Boudol
\thanks{INRIA Sophia-Antipolis}
\and
Fr\'ed\'eric Boussinot
\thanks{INRIA Sophia-Antipolis}
\and
Ilaria Castellani
\thanks{INRIA Sophia-Antipolis}}

\maketitle
\begin{abstract}
In this note we revisit the so-called {\em reactive} programming
style, which evolves from the synchronous programming model of the
\textsc{Esterel} language by weakening the assumption that the absence
of an event can be detected instantaneously. We review some research
directions that have been explored since the emergence of the reactive
model ten years ago. We shall also outline some questions that remain
to be investigated.
\end{abstract}



\section{Introduction}
In synchronous models the computation of a set of participants 
is regulated by a notion of {\em instant}. 
The {\em Synchronous Language} introduced in \cite{BD92}
belongs to this category.  A {\em program} in this language
generally contains sub-programs running in parallel and 
interacting via shared {\em signals}.  
By default, at the beginning of each instant a signal is absent
and once it is emitted it remains in that state till the end
of the instant.
The model can be regarded as a relaxation of the \textsc{Esterel} model
\cite{BG92} where the {\em reaction to the absence} of a signal
is delayed to the following instant, thus avoiding the difficult problems
due to {\em causality cycles} in \textsc{Esterel} programs.

The model has gradually evolved into a
programming language for concurrent applications and has been implemented in
the context of
various programming languages such as \textsc{C}, \textsc{Java},
\textsc{Scheme}, and \textsc{Caml} (see Section~\ref{implementations} below).
The design accommodates a dynamic computing environment with threads
entering or leaving the synchronisation space. In this context,
it seems natural to suppose that the scheduling of the threads is only
determined at run time (as opposed to certain synchronous languages such as
\textsc{Esterel} or \textsc{Lustre}).

The model is based on a {\em cooperative} notion of concurrency.  This
means that by default a running thread cannot be preempted unless it
explicitly decides to return the control to the scheduler. This contrasts
with the model of {\em preemptive} threads, where by default a
running thread can be preempted at any point unless it explicitly
requires that a series of actions is atomic.  We refer to,
e.g.,~\cite{Ous96} for an extended comparison of the cooperative and
preemptive models.  It appears that many typical ``concurrent''
applications such as event-driven controllers, data flow architectures,
graphical user interfaces, simulations, web services, multiplayer
games, are more effectively programmed in a cooperative (and possibly
synchronous) model than in the preemptive one.

The purpose of this note is to revisit the basic model and to review some 
research directions that have been explored since the emergence of the
model ten years ago. We shall also outline some questions that remain
to be investigated.

\section{The basic model}
In this section, we introduce our basic model which is largely inspired by 
the original proposal \cite{BD92}, and, as regards 
concurrency, by the FairThreads model \cite{boussinot:rr-inria-5039}.

We assume a countable set of {\em signal names} $s,s',\ldots$ and we
let \w{Int} be a finite set of signal names representing an observable
{\em interface}. A {\em signal environment} $E$ is a partial function from
signal names to boolean values $\w{true}$ and $\w{false}$ whose domain
of definition $\w{dom}(E)$ is finite and contains \w{Int}. Such an environment 
records the signals that have been emitted during the current instant,
as well as  the ones that exist but are still absent.  The
semantics should preserve the invariant that all signals defined in a
program (see below) belong to the domain of definition of the related
environment.  In particular, all signal names which are not in the
domain of definition of the environment are guaranteed to be fresh,
i.e., not used elsewhere in the program.

We define a {\em thread} as 
an expression written according to the following grammar:
\[\begin{array}{rcl}
T &::=& \nil \Alt (\s{emit}\ s) \Alt 
           (\s{local}\ s\ T) \Alt (\s{thread}\ T)
\\
& \Alt&
           (\s{when} \ s \ T) \Alt (\s{watch} \ s \ T) \Alt A(\vc{s})
           \Alt (T;T)
\end{array}
\] 
where $A(\vc{s}), B(\vc{s}),\ldots$ denote thread identifiers with
parameters $\vc{s}$. As usual, each thread identifier is defined by
exactly one equation $A(\vc{s})=T$.
A thread is executed in the context of a signal environment which is {\em
shared} with other concurrent threads.

The intended semantics is as follows: $\nil$ is the terminated thread;
$(\s{emit}\ s)$ emits $s$, {\em i.e.}\ sets it to $\w{true}$ and
terminates, $(\s{local}\ s\ T)$ creates a fresh signal which is local
to the thread $T$ and executes $T$ (this construct is a binder for the
name $s$ in $T$); $(\s{thread}\ T)$ spawns a thread $T$ which will be
executed in parallel and terminates; $(\s{when} \ s \ T)$ allows the
execution of $T$ whenever the signal $s$ is present and suspends its
execution otherwise; $(\s{watch} \ s \ T)$ allows the execution of $T$
but kills whatever is left of $T$ at the end of the first instant
where the signal $s$ is present, $T;T$ is the usual
sequentialisation. This operational intuition is formalised in 
the following rules, where the predicate $(T,E) \Eval{P} (T',E')$ means
that the thread $T$ in the environment $E$ executes an {\em atomic}
sequence of instructions (possibly none) resulting in the thread $T'$,
the environment $E'$, and the spawning of the {\em multi-set} of
threads $P$. 
\[
(T_1)\  \infer{~}{(\nil,E) \Eval{\emptyset} (\nil,E)}
\]
\[
(T_2)\  \infer{~}{(\s{emit}\ s,E) \Eval{\emptyset} (\nil,E[s:=\w{true}])}  
\]
\[
(T_3)\  \infer{([s'/s]T,E\cup\{s'\mapsto\w{false}\}) \Eval{P} (T',E')
\quad  s'\notin \w{dom}(E)}
{(\s{local}\ s\ T,E) \Eval{P} (T',E')}  
\]
\[
(T_4)\  \infer{~}{(\s{thread}\ T, E)\Eval{\pset{T}} (\nil,E)} 
\]
\[
(T_5)\  \infer{([\vc{s}/\vc{x}]T,E)\Eval{P} (T',E')\quad A(\vc{x})=T}
      {(A(\vc{s}), E)\Eval{P} (T',E')} 
\]
\[
(T_6)\  \infer{E(s)=\w{false}}
      {(\s{when}\ s \ T, E)\Eval{\emptyset} (\s{when}\ s \ T, E)}  
\]
\[
(T_7) \  \infer{E(s)=\w{true}\quad (T,E) \Eval{P} (\nil, E')}
{(\s{when}\ s \ T, E)\Eval{P} (\nil, E')} 
\]
\[(T_8)\  \infer{E(s)=\w{true}\quad (T,E) \Eval{P} (T', E')\quad T'\neq \nil}
{(\s{when}\ s \ T, E)\Eval{P} (\s{when}\ s \ T', E')}
\]
\[
(T_9)\  \infer{(T,E) \Eval{P} (\nil,E')} 
       {(\s{watch}\ s \ T,E) \Eval{P} (\nil,E')} 
\]
\[
(T_{10})\  \infer{(T,E)\Eval{P} (T',E')\quad T'\neq \nil}
       {(\s{watch}\ s \ T,E) \Eval{P} (\s{watch}\ s \ T',E')} 
\]
\[
(T_{11})\  \infer{(T_1,E) \Eval{P_1} (\nil,E_1)\quad 
(T_2,E_1)\Eval{P_2} (T',E')}
{(T_1;T_2,E)\Eval{P_1\union P_2} (T',E')} 
\]
\[
(T_{12})\  \infer{(T_1,E) \Eval{P} (T',E')\quad T'\neq \nil}
{(T_1;T_2,E)\Eval{P} (T';T_2,E')} 
\]
It can be seen from this description of the operational semantics 
that whenever $(T,E) \Eval{P}
(T',E')$ then the execution of $T$ is either {\em terminated}, that is
$T'=\nil$, or {\em suspended}, that is $T'$ is an expression where one
has to execute a subexpression of the form $(\s{when} \ s \ S)$, but
$E'(s)=\w{false}$ (see the rule $T_6$).
In other words, in our cooperative framework, the
$\s{when}$ instruction is the only one that may cause the interruption
of the execution of a thread. 

The implementation of both the $\s{when}$ and the $\s{watch}$
instructions requires a {\em stack}.  For instance, in $(\s{when} \ s_1 \
(\s{when} \ s_2 \ T))$ the computation of $T$ may progress only if
both the signals $s_1$ and $s_2$ are present.  In 
\[
(\s{watch} \ s_1 \
(\s{watch} \ s_2 \ T_1);T_2);T_3,
\]
we start executing $T_1$. Assuming
that at the end of the instant, the execution of $T_1$ is not
completed, the computation in the following instant resumes with $T_3$
if $s_1$ was present at the end of the instant, with $T_2$ if $s_1$
was absent and $s_2$ was present at the end of the instant, and with
the residual of $T_1$, otherwise.  Note that whenever we spawn a new
thread we start its execution with an empty stack of signals, as in the
FairThreads model \cite{boussinot:rr-inria-5039}.

A {\em program} $P$ is a finite non-empty multi-set of threads.  We
denote with $\w{sig}(T)$ (resp.\ $\w{sig}(P)$) the set of signals free
in $T$ (resp.\ in threads in $P$). To execute a program $P$ in an
environment $E$ during one instant, we proceed as follows:
first schedule (non-deterministically) the atomic executions of the threads
that compose it as long as some progress is possible and second 
transform all active $(\s{watch}\ s\ T)$ instructions where
the signal $s$ is present into the terminated thread $\nil$.
To say that a thread $T$ in an environment $E$ is stuck we write 
$(T,E)\ddagger$. This is defined as 
\begin{equation}
(T,E)\ddagger \mbox{ if }(T,E)\Eval{\emptyset}(T,E)
\end{equation}
Notice that if $(T,E)\ddagger$ then $T$ is either terminated or suspended
in the context of $E$.
To perform the {\em abort} operation associated with the $\s{watch}$
construct at the end of the instant, we
rely on the function $\lf \_ \rf_E$ defined as follows:
\[
\begin{array}{c}
\lf P \rf_E = \pset{\lf T \rf_E \mid T\in P} 
\quad 
\lf \nil \rf_E = \nil
\quad
\lf T;T' \rf_E = \lf T \rf_E ; T'  
\\[5pt]
\lf \s{when} \ s \ T \rf_E = 
\left\{
\begin{array}{ll}
(\s{when} \ s \ \lf T \rf_E)   &\mbox{if }E(s)=\w{true} \\
(\s{when} \ s \ T )   &\mbox{otherwise}
\end{array}\right. 
\\[15pt]
\lf \s{watch}\ s \ T \rf_E = 
\left\{ 
\begin{array}{ll}
\nil &\mbox{if } E(s)=\w{true} \\
(\s{watch}\ s \ \lf T \rf_E) &\mbox{otherwise}
\end{array} \right.
\end{array}
\]
We then formalise as follows the execution during an instant of a program 
$P$ in the environment $E$, where we rely on a multi-set notation.
\[
\begin{array}{cc}
(P_1)\ \infer{\qqs{T\in P}{(T,E)\ddagger}}
      {(P,E) \eval (\lf P \rf_E,E)}
&\quad (P_2)\ 
       \infer{\begin{array}{l}
       \xst{T\in P}{\neg (T,E)\ddagger}\quad  (T,E)\Eval{P'} (T',E')\\[3pt]
       (P\minus\pset{T}\union \pset{T'} \union P',E')\eval (P'',E'')
        \end{array}}
      {(P,E)\eval (P'',E'')}
\end{array}
\]
Finally, the input-output behaviour of a program is described by
labelled transitions $P\act{I/O} P'$ where $I,O\subseteq \w{Int}$ are
the signals in the interface which are present at the beginning and at
the end of the instant, respectively. As in Mealy machines, the
transition means that from program (state) $P$ with ``input'' signals
$I$ we move to program (state) $P'$ with ``output'' signals $O$. This is
formalised by the rule:
\[
\begin{array}{c}
(I/O)
\quad 
\infer{(P,E_{I,P}) \eval (P', E') \quad  O=\set{s\in \w{Int} \mid 
E'(s)=\w{true}}}
{P\act{I/O} P'}  
\\ \\
\mbox{where:}
\quad
E_{I,P}(s)= \left\{ 
      \begin{array}{ll}
      \w{true}  &\mbox{if }s\in I \\
      \w{false} &\mbox{if }s\in(\w{Int}\union \w{sig}(P))-I \\
      \w{undefined}     &\mbox{otherwise}
      \end{array} \right. 
\end{array}
\]
Note that we insist on having all free signals of the program
in the domain of definition of the environment.

To conclude this section we give some examples of derived constructions, 
which are frequently used in the programming practice. In what follows 
$(\s{local} \ s_1 \cdots (\s{local} \ s_n\ T)\cdots)$ abbreviates
as $(\s{local}\  s_1,\ldots,s_n\ T)$, and a similar convention is used 
for $\s{when}$ and $\s{watch}$. Moreover, we assume that the signals 
that are introduced in the following encodings (i.e.\ $s$ in $\s{now}$, 
etc.) are fresh, that is they do not occur in the parameters (i.e.\
$s\notin\w{sig}(T)$, etc.).
\[
\begin{array}{rcll}
(\s{await} \ s )  &=& (\s{when} \ s \ \nil) 
\\
(\s{loop} \ T)    &=& A(\vc{s})\quad\mbox{where
  }\set{\vc{s}}=\w{sig}(T),\ A(\vc{s})=T;A(\vc{s}) 
\\
(\s{now} \ T )    &=& (\s{local} \ s \ (\s{emit} \ s);(\s{watch} \ s \ T))    
\\
\s{pause}         &=& (\s{local}\ s \ (\s{now} \ (\s{await} \ s))) 
\\
(\s{exit}\ s)     &=& (\s{emit}\ s); \s{pause} 
\\
(\s{trap}\ s \ T)   &=& (\s{local}\ s \ (\s{watch}\ s\ T)) 
\end{array}
\]
and finally
\[
\begin{array}{rcll}
(\s{present}\ s \ T\ T') &=& 
(\s{local}\ t\ 
(\s{thread}\ (\s{watch}\ s\ \s{pause};(\s{thread}\ T';(\s{emit}\ t)))); \\
&&\phantom{(
\s{local}\ t\ }
(\s{now}\ (\s{await}\ s);(\s{thread}\ T;(\s{emit}\ t))); \\
&&\phantom{(
\s{local}\ t\ }(\s{await}\ t)
)
\end{array}
\]
The instruction $(\s{await} \ s)$ suspends the computation till the
signal $s$ is present.  The instruction $(\s{loop} \ T)$ can be
thought of as $T;T;T;\cdots$. Note that in $(\s{loop} \ T);T'$, $T'$
is {\em dead code}, {\em i.e.}, it can never be executed.  The
instruction $(\s{now} \ T)$ runs $T$ for the current instant, {\em
i.e.}, if the execution of $T$ is not completed within the current
instant then it is terminated.  The instruction $\s{pause}$ suspends
the execution of the thread for the current instant and resumes it in
the following one. We may rely on this instruction to guarantee the
termination of the computation of each thread within an instant.  The
constructs $\s{trap}/\s{exit}$ provide an elementary exception
mechanism.  The instruction $(\s{present}\ s \ T\ T')$ branches on the
presence of a signal. More precisely, if $s$ is emitted during the
current instant, this construction spawns the thread $T$ for
execution, and blocks $T'$ (which is thrown away at the next instant),
while if $s$ is not emitted, the thread $T'$ is executed in the next
instant, and $T$ never gets performed.

\begin{remark}[comparison with \cite{BD92}]
The model we have introduced is largely inspired by the original
proposal \cite{BD92}.  The main novelties or variations are:
replacing parallel composition, the $\s{await}$ and the $\s{loop}$
instructions with, respectively, the \s{thread} and $\s{when}$
constructs, and recursive definitions, and relying on a ``big step''
operational semantics.  We also remark that in the definition of the
conditional branching $(\s{present}\ s \ T\ T')$
the expressions $T$ and $T'$ are under a
\s{thread} instruction.  This implies that their execution does {\em
not} depend on when or watch signals that may be on top of them. If
this must be the case, then we may prefix $T$ and $T'$ with
suitable \s{when} and \s{watch} instructions.
\end{remark}

\section{Implementations and applications}
\label{implementations}
Several implementations related to the model described in the previous
section have been proposed over the years. Here, we briefly review
some of them (in a more or less chronological order), highlighting
their main features.

Reactive-C \cite{boussinot:rc91} was proposed as a preprocessor of C
for assembly-like reactive programming, and it has been used to
implement SL.  There also exists a reactive library very close to
Reactive-C written in Standard ML \cite{pucella:reactive}. Two sets of
Java classes have been designed for reactive programming in Java:
SugarCubes \cite{fb-jfs:sugar98} and Junior \cite{Junior}. In these
implementations, reactive threads are not mapped on Java threads and
thus the problems raised by the latter (for example, the limitation on
their number or their memory footprints) are avoided.  Icobjs
\cite{brunette:oopsla02} is a framework for graphical reactive
programming, built on top of SugarCubes. Icobjs have been used for
video games, simulations in physics and simulations of the Ambient
calculus.  Both Java and ML have been extended with reactive
primitives, respectively in Rejo \cite{acosta:notere00} and ReactiveML
\cite{MandelPouzetPPDP05}.  FairThreads
\cite{boussinot:rr-inria-5039,SchemeFT} and Loft \cite{site:Loft}
define a thread-based framework in which reactive cooperative threads
and preemptive threads can be used jointly.  Finally, ULM
\cite{Boudol04,Epardaud} proposes to use reactive programming,
enriched with migration primitives, for global computing over the Web.
This takes advantage of the fact that reactive programming, as opposed to
the synchronous model of \textsc{Esterel} for instance, is well-suited
for applications involving dynamic concurrency.

Starting from the work initiated by Laurent Hazard on Junior, a lot of
effort has been devoted to designing efficient implementations of
reactive frameworks. Efficiency mainly comes from the absence of
busy-waiting of suspended threads waiting for an event, and from
scheduling techniques allowing direct access to the next thread to
execute. As examples of efficiency-critical applications recently
implemented using the reactive style, we may mention the simulation of
a complex network routing protocol for mobile ad-hoc networks described
in ReactiveML \cite{MandelPouzetPPDP05}, the implementation of a Web
server in Scheme \cite{SchemeFT}, and the implementation of cellular
automata in \cite{Boussinot04}, which we shall now describe in some details.

Cellular automata (CA) are used in various simulation contexts, for
example, physical simulations, fire propagation, or artificial
life. These simulations basically consider large numbers of
small-sized identical components, called cells, with local
interactions and a global synchronized evolution. Conceptually, the
evolution of a CA is decomposed into couples of steps: during the
first step, cells get information about the states of their neighbours
and during the second step they change their own state according to
the information obtained from the previous step. Usually, CA are coded
as sequential programs, basically made of a single main loop which
considers all cells in turn. Using the reactive style to program
cellular automata, where each cell is a reactive thread, has the following 
advantages:
\begin{itemize}
\item
Instants naturally represent steps: at each instant, each cell changes
its state according to the neighbours states at the previous instant,
signals its new state, and then waits for the information about the state
of its neighbours.
\item 
The behaviour of cells coded as look-up tables in usual CA
implementations is rather opaque. This is generally not felt as a big
issue because cells behaviours are often very simple. However, in some
contexts, for example artificial life, one may ask for more complex
cell behaviours. In these cases, the modularity obtained with reactive
programming is an advantage.
\item
One can obtain efficient implementations of CA spaces in which each cell
is implemented as a thread. To improve efficiency, cells can be
created only when needed. Note that quiescent cells (with no active
neighbour) are just waiting for an activation signal; their presence
thus does not introduce any overhead at execution.
\end{itemize}

Reactive programming focusses on behaviours rather than on
data. Entities found in video games can thus be naturally coded using
reactive primitives. Similarly, we have also used the reactive model
for interactive simulation of physical systems. Indeed, the reactive
style provides us with a very simple and modular way to describe the
evolution of complex physical systems. The main features of this 
approach are simplicity of model
construction and high modularity of components. This approach allows us
to express both continuous and discrete aspects of a model. For
example, consider a planet/meteor system. A planet is implemented with a
behaviour which, at each instant, emits a gravity signal with its
coordinates.  A meteor, at each instant, waits for the gravity signal
and moves accordingly. One thus gets systems made of interacting
components in which new components can be dynamically added. Applets
illustrating this approach, coded in SugarCubes, are available on
the Web \cite{mimosarp}.

\section{Some issues}
In this section we briefly discuss some issues related to reactive 
programming.

\subsection{Values}\label{values}
Practical programming languages that have been developed on top of the
basic reactive model include {\em data types} beyond pure signals. For
instance, we may have the inductive type of booleans $\w{bool} = \s{t}
\mid \s{f}$, and the inductive type of natural numbers in unary
notation $\w{nat} = \s{z} \mid \s{s} \ \w{of} \ \w{nat}$. At the very
least, the reactive kernel embedded in a general purpose language
should include ways of using the values manipulated in this language.
There are
two main approaches to adding values to the model: (1) to introduce
references as in the \textsc{ML} language, and (2) to assume that signals
carry values and that the last emission ``covers'' in a sense the
previous ones (if any).  In the latter case, an important design
choice to make is to decide what is ``the'' value associated with a
signal at a given instant, and what is the corresponding construct for
consulting this value. The simplest model is to regard the value of a
signal as ephemeral. That is, the value is updated, as for a
reference, by the next emission of the given signal. However, this is
not quite compatible with the idea that a signal is broadcast, and
that all the running threads have a consistent view of it -- either
present or absent -- at each instant. Therefore, some other mechanisms
have been designed. In \textsc{Esterel} for instance, one assumes for
each type of signal value a function for combining the various values
emitted on that signal, and the actual value carried by the signal at
some instant is the combination of all the values emitted during this
instant (in \textsc{Esterel}, with the strong synchrony hypothesis,
the combination function should be associative and commutative, since
the result should be independent of any scheduling). A
similar approach has been followed in SugarCubes \cite{fb-jfs:sugar98}
and ReactiveML \cite{MandelPouzetPPDP05}. Notice that in the 
reactive model, where one cannot statically predict that a signal will
or will not be emitted, one has to collect the value of a signal only
at the beginning of the next instant. One may also trigger a processing
mechanism each time a value is emitted on a signal.
Another possibility that is considered in some implementations is to
specify, in a receive statement, the rank of the value (in the
emission order) in which one is interested.

\subsection{Reactivity}\label{reactive}
A first property that we would like to ensure regarding reactive 
programs is that they should indeed be reactive, in the following 
(coinductive) sense:

\begin{definition}
A program $P$ is {\em reactive} if for every choice $I$ of the
input signals there are $O,P'$ such that $P\act{I/O}P'$ and $P'$ is 
reactive.
\end{definition}
The reactivity property is not for free. For instance, the thread
$A=(\s{await}\ s ); A$ may potentially loop within an instant.
Whenever a thread loops within an instant the computation of the whole
program is blocked as the instant never terminates.  
One approach to ensure reactivity is to produce a static
analysis that guarantees that all loops that may occur within an 
instant traverse a \s{pause} instruction.

While reactivity is a necessary property, it does not guarantee that
in practice the program will react for arbitrarily many instants and
that this will happen within reasonable time and/or space.  A first
problem has to do with the implementation of the \s{when} and
\s{watch} instructions.  Consider, the thread $A= (\s{local} \ s \
(\s{watch} \ s \ \s{pause}; A))$.  Every time the execution crosses
the \s{watch} instruction it causes the insertion of a new signal $s$
which may potentially abort the execution (although this is not the case
with this particular program). Thus the execution of this
program may potentially cause a stack overflow. This kind of
pathological programs can be removed by a static analysis that checks
that there is no loop in the program (possibly going through several
instants) that may cause an increase of the stack.

A second problem is due to the fact that the number of (active)
threads and signals may grow without limit. Indeed, it can be shown that 
our basic language is Turing complete. In
practice, we need to control the number of threads, and in this respect
an interesting feature of the language is the \s{watch} instruction
which allows to terminate explicitly the execution of a thread (at the
end of an instant).

Finally, a third problem, as regards reactivity, is caused by the
introduction of data values.
The size of the values we are interested in, like lists or trees, is
usually not a priori bounded. What does it mean to ensure reactivity
in this case?  We have in \cite{AD04,AD05} considered three
increasingly ambitious goals in this respect. A first one
is to ensure that every instant terminates. A second one is to
guarantee that the computation of an instant terminates within
feasible bounds which depend on the size of the parameters of the
program at the beginning of the instant. A third one is to
guarantee that the parameters of the program stay within certain
bounds, and thus the resources needed for the execution of the system
are controlled for arbitrarily many instants.  In particular, we have
been adapting and extending techniques developed in the framework of
(first-order) functional languages.  The general idea is that
polynomial time or space bounds can be obtained by combining
traditional termination techniques for term rewriting systems with an
analysis of the size of computed values based on the notion of
quasi-interpretation (\cite{Amadio04,BMM01}).  Thus,
in a nutshell, ensuring ``feasible reactivity'' requires a
suitable termination proof and bounds on data size.

\subsection{Determinism}\label{det}
We say that two programs $P,P'$ are equal up to renaming
if there is a bijection from $\w{sig}(P)$ to $\w{sig}(P')$
that is the identity on the observable signal names in the
interface $\w{Int}$ and that when applied to $P$ produces $P'$.
As usual, an inspection of the semantics shows that the 
observable behaviour of a program does not depend on the 
specific choice of its internal signal names. First we define 
deterministic programs. As with the notion of reactivity, determinism
should hold at every instant, and therefore our definition is coinductive.

\begin{definition}
A program $P$ is {\em deterministic} if for every choice $I$ of the
input signals if $P\act{I/O_{1}} P_1$ and $P\act{I/O_{2}} P_2$
then $O_1 = O_2$ and $P_1 = P_2$ up to the same renaming, and $P_1$
is deterministic.
\end{definition}
It is immediate to verify that the evaluation of a thread $T$ in an
environment $E$ is deterministic.  Therefore the only potential source
of non-determinism comes from the scheduling of the threads.  The
basic remark is that the emission of a signal can
never block the execution of a statement within an instant. The more
we add signals the more the computation of a thread can progress
within an instant. Of course, this property relies on the fact that we
cannot detect the absence of a signal before the end of the instant.

\begin{proposition}\label{deterministic-prop}
All programs are deterministic.
\end{proposition}
Clearly, this property is likely to be lost when adding values to the
model. Assuming that we have valued signals, consider for instance the
program $P = \pset{(\s{emit} \ s \ \s{t}), (\s{emit} \ s
\ \s{f})}$ where two threads emit the boolean values \s{t} and \s{f},
respectively, on the signal $s$.  The value which is observed on the
signal at the end of the instant depends on the scheduling of the
threads (unless the values are combined using an associative and
commutative function, as in \textsc{Esterel}). So it seems that we
have to accept the idea that when introducing data types the result of
the program depends on the scheduler.  In practice, one may assume
that the scheduler is {\em deterministic} in the program and the
input.  This is a significant difference with preemptive concurrency.
In preemptive concurrency, the scheduling policy may depend on factors
such as the current workload which are {\em independent} from the
program and the input.  Assuming a deterministic scheduler has a
positive effect on the process of testing, tracing, and debugging
concurrent programs.  Besides determinism, it might be reasonable to
put additional constraints on the scheduler.  One such constraint is
the following: if a thread suspends its execution during an instant
then all the threads that are ready to run at the moment of the
suspension will be given a chance to progress before the computation
of the suspended thread is resumed (if ever). With such a scheduler in
mind, it makes sense to define:
\[
\s{yield} = (\s{local} \ s \ (\s{thread}\ (\s{emit} \ s));(\s{await} \ s ))
\]

\subsection{Program equivalence}\label{prog-equivalence}
We have described the operational semantics of reactive and deterministic 
programs as a reaction to a given input, producing a unique output and 
continuation. Looking for a more abstract, extensional semantics,
one possibility is to consider that it is determined by
the set $\w{tr}(P)$ of infinite traces associated with the possible
runs of the program $P$. Namely:
\[
\w{tr}(P) = \set{ (I_1/O_1)(I_2/O_2)\cdots \mid 
                  P\act{I_{1}/O_{1}} P_1 \act{I_{2}/O_{2}} P_2 \cdots}
\]
Another possibility could be to define a notion of
bisimulation. Namely, consider the largest (symmetric)
relation $R$ on programs that satisfies the following
condition:  for every  $(P,P')\in R$  and input $I$, 
if $P\act{I/O}P_1$ then  $P'\act{I/O} P'_{1}$ and 
$(P_1,P'_{1})\in R$. It is important to notice that
for our deterministic language these two notions coincide.

\begin{proposition}\label{trace-bisimulation}
Two reactive and deterministic 
programs are trace equivalent iff they are bisimilar.
\end{proposition}
Of course, this reduces considerably the debate on what the right
notion of program equivalence is.  The notion of weak bisimulation --
another familiar concept in the semantics of concurrency -- is also
missing.  However, we must point out that, although the problem of
defining program equivalence has an obvious solution, little work has
been done so far on the problem of defining and characterising a
suitable notion of {\em thread equivalence} which is preserved by
program contexts. Moreover, as we have seen, adding values to the
language turns it into a non-deterministic model, for which no notion
of equivalence has been investigated so far.

\end{document}